\documentstyle[prb,aps,epsfig,twocolumn]{revtex}

\begin{document}

\title{On the magnetic stability at the surface in strongly correlated electron
   systems}
\author{T. Herrmann and W. Nolting}
\address{Institut f\"ur Physik, Humboldt-Universit\"at zu Berlin,                
           Invalidenstr.\ 110, 10115 Berlin, Germany}
\maketitle
\begin{abstract}
The stability of ferromagnetism at the surface at finite temperatures is
investigated within the strongly correlated Hubbard model on a semi-infinite
lattice. 
Due to the reduced surface coordination number 
the effective Coulomb correlation
is enhanced at the surface compared to the bulk. 
Therefore, within the well-known Stoner-picture of band ferromagnetism
one would expect the magnetic stability at the surface to be enhanced as well.
However, by taking electron correlations into account well 
beyond the Hartree-Fock (Stoner)
level we find the opposite behavior:
As a function of temperature
the magnetization of the surface layer decreases
faster than in the bulk. 
By varying the hopping integral within the surface layer this behavior becomes 
even more pronounced. A reduced hopping integral at the surface tends to destabilize
surface ferromagnetism whereas the magnetic stability gets enhanced by an
increased hopping integral. This behavior represents a pure correlation effect
and can be understood in terms of general arguments which are based
on exact results in the
limit of strong Coulomb interaction.  
\end{abstract}


\section{Introduction}
There is currently active interest in the 
influence of reduced translational
symmetry on the magnetic properties at surfaces, in ultra-thin films and
in multi-layer structures.  One class of materials that is intensively studied are 
the magnetic transition metals, within which the magnetically active electrons are
itinerant. 
Theoretically  it is possible to calculate the ground state properties like the
magnetic moments at and near surfaces and in thin films in great detail 
by use of 
{\em ab initio} methods \cite{FF87,KMS94}. However, these approaches are strictly based on a
Stoner-type picture of ferromagnetism and treat electron correlation
effects that are responsible for the spontaneous magnetic order on a low level
only. Further, an extension to finite temperatures  states a difficult problem. 

To study the temperature behavior of spontaneous 
ferromagnetism rather idealized model systems have proven to be a good starting point.   
What concerns the magnetic properties of systems with reduced translational
symmetry, the temperature behavior has been, up to now, almost exclusively
studied within localized spin models (Ising model, Heisenberg model) 
\cite{HBC+72,HLN79,EM91a,JDB92,SY92,SN99}.
In these systems the magnetic stability at the surface is found to be reduced 
compared to the bulk. The reduced magnetic stability  results simply from  
the reduced exchange coupling at the surface due 
to the lower coordination number. 
However, it is not clear at all to what extent
this argument applies also for the magnetic transition metals Fe, Co, Ni, 
which are prototypical materials 
of  the so-called bandmagnetism of itinerant electrons.
As it is well known, within a tight-binding scheme for the description of the
bandstructure the reduced coordination
number at the surface leads to a reduced effective bandwidth in the surface
layer. As a consequence the effective correlation at the surface will be
enhanced. Thus, on the basis of the well-known Stoner-picture of bandmagnetism
one may intuitively expect the magnetic stability at the surface to be enhanced
compared to the bulk. 

In this paper we investigate the  magnetic stability at the surface in strongly
correlated electron systems within the 
Hubbard model on a semi-infinite lattice.
Based on the limit of strong Coulomb interaction we give general arguments
concerning the magnetic stability at the surface. 
In addition we will show results of numerical
evaluations within the spectral density approach to the semi-infinite Hubbard
model. The spectral density approach is based on exact results about the general
shape of the spectral density in the limit of strong Coulomb interaction. 
To emphasize the importance of correlation effects we will show, in addition,
results obtained within the Hartree-Fock approximation (Stoner-model).

\section{Theory}
To describe the geometry of the semi-infinite lattice each lattice vector is
decomposed into two parts ${\bbox R}_{i\alpha}={\bbox R}_i+{\bbox r}_\alpha$.
Throughout the paper
${\bbox R}_i$ (Latin index) denotes a lattice vector parallel
to the film surface whereas ${\bbox r}_\alpha$ (Greek index) 
refers to the distance from the surface ($\alpha=1$: surface layer).
Each lattice plane parallel to the surface contains $N$ lattice sites. 
Within the layers we assume translational invariance. 

Using this notation 
the Hubbard model for the semi-infinite lattice reads:
\begin{equation} \label{hub_op}
        {\cal H}=\sum_{i,j,\alpha,\beta,\sigma}
	(T_{ij}^{\alpha\beta}-\mu\delta_{ij}^{\alpha\beta})
	c_{i\alpha\sigma}^{\dagger}c_{j\beta\sigma}+
	\frac{U}{2}\sum_{i,\alpha,\sigma}n_{i\alpha\sigma} n_{i\alpha-\sigma}.
\end{equation}
Here $c_{i\alpha\sigma}$ ($c_{i\alpha\sigma}^{\dagger}$) stands for 
the annihilation (creation) operator of an electron with spin $\sigma$ 
at the lattice site ${\bbox R}_{i\alpha}$,
$n_{i\alpha\sigma}=c_{i\alpha\sigma}^{\dagger}c_{i\alpha\sigma}$ 
is the number operator. 
$U$ denotes the on-site Coulomb matrix element and $\mu$ 
the chemical potential.
$T_{ij}^{\alpha\beta}$ is the hopping integral
between the lattice sites ${\bbox R}_{i\alpha}$ and ${\bbox R}_{j\beta}$. 

Let us first discuss the kinetic part of the  system ($U=0$) that is treated
within the tight-binding approximation. 
A two-dimensional Fourier transformation yields the corresponding dispersions
\begin{eqnarray}
\label{hopping}
   T_{\bbox k}^{\alpha\beta}&=&\frac{1}{N}\sum_{ij} T_{ij}^{\alpha\beta} 
                            e^{-i{\bbox k}({\bbox R}_{i}-{\bbox R}_{j})}\\
			  &=& T_{0\alpha}\delta^{\alpha\beta}+
			      t\gamma^{\alpha\beta}({\bbox k}), 
\end{eqnarray}
that can easily be calculated.
Here and in the following ${\bbox k}$ denotes a wave-vector from the
underlying two-dimensional (surface) Brillouin zone.
$t$ is the hopping integral between nearest neighbor atoms.
The on-site hopping integral $T_{0\alpha}=T_{ii}^{\alpha\alpha}$ 
gives the center of gravity of the dispersion of the 
$\alpha$-th layer.
The dispersions $T_{{\bbox k}}^{\alpha\beta}$ determine the Bloch density of
states (BDOS) $\rho_{0\alpha}(E)$.
The bandwidth of the BDOS is denoted by $W$. 
A moment analysis of the BDOS  
($\Delta_{\alpha}^{(n)}=\int dE \,E^n \rho_{0\alpha}(E)$)
shows the variance at the surface  $\Delta_{s}^{(2)}$ ($\alpha=1$) 
to be considerably reduced compared to the variance of the inner
layers $\Delta_{b}^{(2)}$ ($\alpha=2,\,\dots,\,\infty$).
Mathematically, the square root of the variance 
($\mbox{\scriptsize$\sqrt{\mbox{$\Delta_\alpha^{(2)}$}}$}$) 
gives a measure of the effective bandwidth of the BDOS. 
In addition, for an asymmetric BDOS also the skewness 
$\Delta_{\alpha}^{(3)}$ is reduced at the surface compared to the bulk. 
For fcc-type semi-infinite lattices with  surface orientations (100) and (111)
that are considered in the numerical evaluation (see below)
we find:
\begin{equation}\label{eq:mom_bdos}
\begin{array}{ll}
(\Delta^{(2)}_s/\Delta^{(2)}_b)^{(100)} =  0.667, & 
(\Delta^{(2)}_s/\Delta^{(2)}_b)^{(111)} = 0.750,\\
(\Delta^{(3)}_s/\Delta^{(3)}_b)^{(100)} =  0.5, & 
(\Delta^{(3)}_s/\Delta^{(3)}_b)^{(111)} = 0.625.\\
\end{array}
\end{equation}
The corresponding surface coordination numbers are given by 
$z^{(100)}_s=8$ and $z^{(111)}_s=9$. These have to be compared to the bulk
coordination number of the fcc lattice   $z^{(fcc)}_b=12$.

The Hubbard model states a complex  
many-body problem, the rigorous solution of which is possible up to now
only for some limiting cases \cite{LW68,MV89,GKKR96}. In thin films and semi-infinite systems even more
complications are introduced due to the reduced translational symmetry at the
surface. 
The standard Hartree-Fock approximation (HFA), which has been applied
previously for the investigation of thin film ferromagnetism \cite{GM94,PM95},
is known to drastically
overestimate the magnetic region in the phase diagram.
This is clearly due to the fact that electron correlations are 
treated on a low level  within the HFA. 
Since ferromagnetism in the Hubbard model is an effect of strong Coulomb
interaction we require an approximation scheme which takes correlation effects
into account more reasonably. On the other hand, it must be simple enough to
allow for a detailed study  of phase transitions in semi-infinite systems. 

An approximation scheme which has proven to include 
the essential physics of
spontaneous ferromagnetism in the Hubbard model is the spectral density approach
(SDA) \cite{NB89,HN97a}. 
The SDA can be interpreted as a self-consistent 
extension of the $W/U$ perturbation theory of Harris and Lange \cite{HL67}
in a straightforward way
to intermediate coupling strength as well as finite temperatures.
Despite its rather simple  concept, the SDA leads to convincing results   
with regard to the qualitative behavior of the ferromagnetic 
solutions of the Hubbard model.
A similar approach has been applied to a multiband Hubbard model with accurate
results for the magnetic key quantities of the prototype ferromagnets Fe, Co, and
Ni \cite{NBD+89,VN96,NVF95}. 
Recently a generalization of the SDA
has been proposed to deal with the modifications due to the reduced
translational symmetry \cite{PN96,HPN98,Her99}. 
In the following we give only a brief derivation of the SDA for a semi-infinite
lattice. For further details we refer the reader to previous publications 
\cite{NB89,HN97a,PN96,HPN98,Her99}.

The basic quantity from which all relevant information on the system can be
obtained 
is the retarded single-electron 
Green function 
\begin{equation}\label{green}
       G_{ij\sigma}^{\alpha\beta}(E)=\langle\langle c_{i\alpha\sigma};
       c_{j\beta\sigma}^{\dagger}\rangle\rangle_E.
\end{equation}
After a two-dimensional Fourier transformation 
one obtains from $G_{ij\sigma}^{\alpha\beta}(E)$ 
the spectral density
$   S_{{\bbox k}\sigma}^{\alpha\beta}(E)=-\frac{1}{\pi} 
   \textrm{Im} G_{{\bbox k}\sigma}^{\alpha\beta}(E)$,
which represents the bare line-shape of a (direct, inverse) photoemission
experiment.  
The diagonal elements of the Green function determine the 
spin- and layer-dependent quasiparticle density of states (QDOS):
        $\rho_{\alpha\sigma}(E)
        =-\frac{1}{\pi} \textrm{Im}G_{ii\sigma}^{\alpha\alpha}(E-\mu).$
Via an energy integration one immediately gets from 
$\rho_{\alpha\sigma}(E)$ the band occupations
\begin{equation}\label{nas}
   n_{\alpha\sigma}\equiv\langle n_{i\alpha\sigma}\rangle=
   \int\limits_{-\infty}^{\infty} dE f_{-}(E) \rho_{\alpha\sigma}(E).
\end{equation}
$\langle\dots\rangle$ denotes the grand-canonical average and 
$f_{-}(E)$ is the Fermi function.
Ferromagnetism is indicated by a spin-asymmetry in the band occupations
$n_{\alpha\sigma}$ leading to non-zero  magnetizations 
$m_\alpha=n_{\alpha\uparrow}-n_{\alpha\downarrow}$.
The mean band occupation $n$ and the mean magnetization $m$ are given by
$n=\frac{1}{d}\sum_{\alpha\sigma}n_{\alpha\sigma}$ and
$m=\frac{1}{d}\sum_{\alpha}m_\alpha$, which are, of course, identical to the
respective bulk values ($\alpha\rightarrow\infty$).

For the generalization of the SDA to systems with reduced translational symmetry
we adopt the local approximation for the self-energy
($\Sigma_{ij\sigma}^{\alpha\beta}(E)=\delta_{ij}\delta^{\alpha\beta}
\Sigma_{\alpha\sigma}(E)$) which has been tested
recently for a semi-infinite lattice \cite{PN97c}. 
The decisive step is to find a reasonable ansatz for the self-energy 
$\Sigma_{\alpha\sigma}(E)$. 
Guided by the exactly solvable atomic limit 
of vanishing hopping ($t=0,\,W=0$)  
and  by the findings of Harris and Lange \cite{HL67} in the strong-coupling 
limit ($U/W\gg 1$), 
a one-pole ansatz for the self-energy $\Sigma^{\alpha}_{\sigma}(E)$ can be
motivated \cite{PN96,Her99}: 
\begin{equation}
\label{eq:one_pole}
\Sigma^{\alpha}_{\sigma}(E)=g^{\alpha}_{1\sigma}
\frac{E-g^{\alpha}_{2\sigma}}{E-g^{\alpha}_{3\sigma}}.
\end{equation}
The spin- and layer-dependent parameters $g^{\alpha}_{1\sigma}$,
$g^{\alpha}_{2\sigma}$ and $g^{\alpha}_{3\sigma}$
are fixed  by exploiting the equality
between two alternative but exact representations
for the moments of the 
Green function:
\begin{eqnarray}
 &&\hspace{-5ex}
   -\frac{1}{\pi}\textrm{Im}
      \int\limits_{-\infty}^{\infty}\! dE\, E^{m} 
	\,G_{ij\sigma}^{\alpha\beta}(E)=\label{mom1}\\
 &&\hspace{10ex}
 \bigg\langle\!\Big[
                       \underbrace{\big[...
	               [c_{i\alpha\sigma},{\cal H}]_{\!-}... ,  
                       {\cal H}\big]_{\!-}}_{m-\textrm{times}},
                        c_{j\beta\sigma}^{\dagger}
                  \Big]_{+}\!\bigg\rangle.\nonumber
\end{eqnarray}

Here $[\dots,\dots ]_{-(+)}$ denotes the commutator (anticommutator).
Eq. (\ref{mom1})  imposes rigorous sum rules on the Green
function  which have been
recognized to state important guidelines 
when constructing approximate solutions for
the Hubbard model \cite{PHW+98}.
By comparing various approximation schemes it has been shown that the inclusion
of the first four sum rules is vital for a reasonable  
description of spontaneous
ferromagnetism in the Hubbard model, especially at finite temperatures
\cite{PHW+98}.
Taking the first four sum rules ($m=0,\dots,3$) into account to determine the 
free parameters of
the SDA ansatz for the self-energy 
yields a closed set 
of equations\cite{PN96,HPN98,Her99} that has to be solved self-consistently.
Note, that the self-energy incorporates the so-called 
bandshift $B_{\alpha-\sigma}$ which is layer-dependent and
consists of higher correlation functions\cite{NB89,HN97a,PN96,HPN98,Her99}.
$B_{\alpha-\sigma}$
comes  into play through the fourth sum rule ($m=3$). 
Fortunately, the bandshift can exactly be expressed by the self-energy and the
Green function\cite{NB89,HN97a,PN96,HPN98,Her99}. 
A spin-asymmetry of the bandshift, which can occur 
within the  self-consistent calculation
in certain parameter regions, may generate 
or stabilize ferromagnetic solutions in the strongly 
correlated Hubbard model.
For strong Coulomb interaction the bandshift  
is found\cite{HL67,HN97a,PHW+98,Her99}
to be proportional to the
kinetic energy of the system 
($n_{\alpha-\sigma}(1-n_{\alpha-\sigma})B_{\alpha-\sigma}\sim E_{kin,-\sigma}$,
$U\gg W$).

\section{Results and discussion}
The numerical evaluations have been done for a semi-infinite fcc lattice. To study
the influence of the reduced coordination number we have considered two
different surface orientations (100) and (111). The mean band occupation 
is kept fixed at $n=1.4$ and the Coulomb interaction is chosen 
to be $U=3W$ which
clearly refers to the strong coupling region.
In order to avoid a mixture of charge transfer and electron correlation effects we
have  excluded charge transfer between the layers by requiring 
$n_\alpha=n,\,\,\forall\alpha$. 
This is achieved by
adjusting the layer-dependent on-site hopping integral $T_{0\alpha}$.

Fig.~\ref{fig_1} shows the spontaneous magnetization of the semi-infinite fcc
lattice as a function of increasing temperature. At low temperatures the system
shows full polarization ($m=2-n$) 
and is uniformly
magnetized ($m_\alpha=m,\,\,\forall\alpha$) since we have explicitly excluded
charge transfer between the layers.
As the temperature increases the magnetization of the
surface layer $m_1$ decreases faster than the  bulk magnetization $m_b$.
This behavior holds for both considered surface orientations, (100) and (111).
The deviation  from the bulk magnetization
is almost completely restricted to the topmost layer.
Note, that the influence of the surface 
is stronger for the more open (100) structure.
Due to the coupling between the layers 
that is induced by the electron hopping a
unique Curie temperature is found for the whole system. 
We have to conclude from Fig.~\ref{fig_1}, that 
within the strongly correlated
Hubbard model the magnetic order is less stable at the surface than 
in the bulk, although the effective correlation is enhanced at the surface.
This observation clearly contradicts the expectation 
on the basis of the well known Stoner-picture
of band ferromagnetism. Indeed, repeating the same calculations within the
 HFA we find the magnetic order  
to be more stable at the surface compared to the bulk. 
For very strong Coulomb
interaction the layer magnetizations within the HFA become almost independent of
the layer index.

The reduced magnetic stability at the surface in strongly correlated electron
systems  can be 
understood in terms of general arguments
by use of the moment analysis of the BDOS of the semi-infinite
lattice (see Eq.~(\ref{eq:mom_bdos})).
The arguments are based on the magnetic behavior of bulk systems with full
translational symmetry, 
which have been studied in very detail\cite{PHW+98,HUM97,HN97b,Ulm98,WBS+98}.  
Both the reduced variance 
($\Delta_{s}^{(2)}<\Delta_{b}^{(2)}$) as well as the reduced skewness 
($\Delta_{s}^{(3)}<\Delta_{b}^{(3)}$) at the surface lead to the
same trend:

\begin{figure}
 \centerline{\epsfig{file=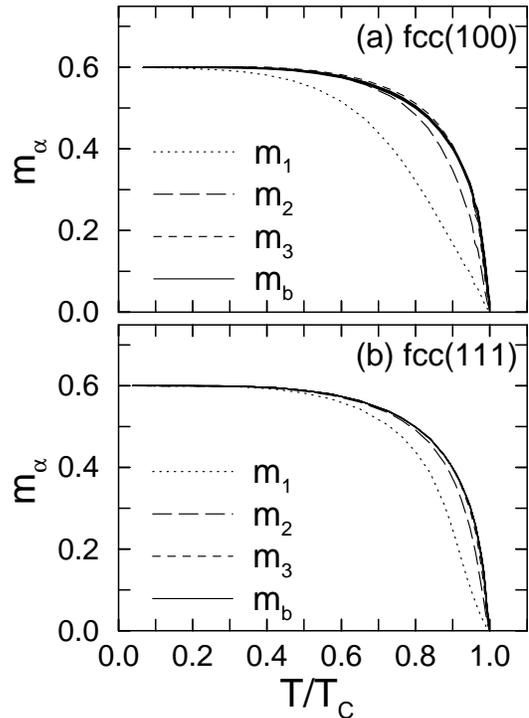,width=0.8\linewidth}}
\caption{Magnetization $m_\alpha$ 
for a semi-infinite fcc lattice as a function of the reduced 
temperature $T/T_C$. $\alpha=1$ corresponds to the surface layer.
(a) (100) surface; (b) (111) surface. 
Further parameters: $U/W=3$; $n=1.4$} \label{fig_1}
\end{figure}

(i) Within various approximation schemes 
\cite{HUM97,HN97b,PHW+98} as well as DMFT-QMC 
calculations \cite{Ulm98,WBS+98}
it has been established, that a large skewness of the BDOS tends to stabilize
spontaneous ferromagnetism in the Hubbard model.  Since the skewness of the
surface layer BDOS is reduced compared to the inner layers this explains the
trend to a reduced magnetic stability at the surface.

\begin{figure}
 \centerline{\epsfig{file=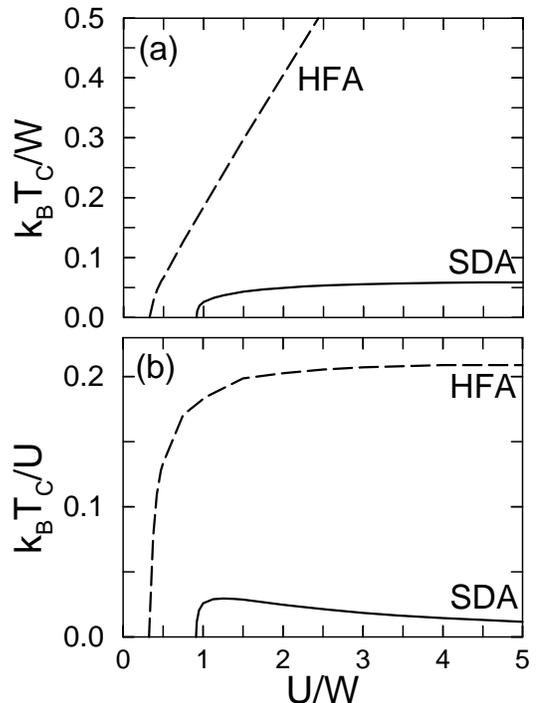,width=0.8\linewidth}}
\caption{Curie temperature $T_C$ as a function of the effective correlation
$U/W$ for an fcc lattice (bulk calculation). 
(a) constant bandwidth $W$ of the BDOS . (b) constant Coulomb
interaction $U$. $k_B$ is the Boltzmann constant.
Further parameters: $n=1.4$
} \label{fig_2}
\end{figure}

(ii) In order to understand the influence 
of the reduced variance of the BDOS at the surface
on the magnetic stability  
we have plotted in Fig.~\ref{fig_2} the
Curie temperature $T_C$ for an fcc bulk system in two different ways. 
Fig.~\ref{fig_2}(a) shows $k_BT_C/W$ as a function of the effective correlation
$U/W$. This corresponds to the situation where the bandwidth $W$ of the
uncorrelated Bloch band is kept fixed while the  
Coulomb interaction $U$ is varied.
Within the SDA the Curie temperature  steeply increases as soon as $U/W$ exceeds
a critical lower bound.  But already for  $U/W\approx 1-2$  
the Curie temperature starts to saturate and 
approaches a finite  
value as $U\rightarrow\infty$.
This behavior can be understood by inspecting the quasiparticle spectrum.
In the strongly correlated Hubbard model the quasiparticle  spectrum 
splits into a high and a low energy subband
(''Hubbard-bands''; for a detailed discussion see, e. g., Refs. \cite{HN97a,HPN98}). 
Both subbands are separated by an energy amount of approximately $U$.
In the ferromagnetic phase
there is an additional spin splitting 
within each subband between the majority and the minority spin
direction. According to the 
results of Harris and Lange \cite{HL67},  the spin splitting 
is -- for large $U$ -- determined by
the spin-asymmetry in the kinetic 
energy $E_{kin}=E_{kin,\uparrow}+E_{kin,\downarrow}$ 
of the system \cite{HL67,HN97a,PHW+98,Her99}.
This can be seen, e. g., for the spin-splitting $\Delta_{2,ex}$
between the centers of gravity
($T_{2\sigma}$) of the upper Hubbard-bands  
within which the chemical potential $\mu$
is located for the more than half-filled case ($n>1$). 
Here one finds\cite{HL67,HN97a,PHW+98,Her99}:
\begin{eqnarray}
\Delta_{2,ex}&=&T_{2\downarrow}-T_{2\uparrow}\nonumber\\
&\mbox{\raisebox{-0.5ex}{$\stackrel{U\gg W}{\longrightarrow}$}}&
(1-n_{\uparrow})B_{\uparrow}-(1-n_{\downarrow})B_{\downarrow}\nonumber\\
&\mbox{\raisebox{-0.5ex}{$\stackrel{U\gg W}{\longrightarrow}$}}&
\frac{E_{kin,\uparrow}}{n_{\uparrow}}-\frac{E_{kin,\downarrow}}{n_{\downarrow}}
\label{eq:splitting}
\end{eqnarray}
The kinetic energy, however,  becomes 
independent of $U$ in the limit of 
strong Coulomb interaction and the bandwidth $W$ remains the only 
relevant energy scale with respect to the magnetic properties of
the system.
Therefore, the spin splitting (\ref{eq:splitting}) 
as well as the Curie temperature
saturate as a function of increasing $U$ (Fig.~\ref{fig_2}(a)).
The same behavior  has been observed within several strong coupling approaches
to the Hubbard model \cite{HN96,PHW+98} 
as well as DMFT-QMC calculations \cite{WBS+98}. 
Within the HFA the spectrum shows a completely different
behavior. Here, the Hubbard splitting is absent and the spin splitting is given
by $Um$. Both, the  spin splitting and  the Curie temperature  increase linearly
with $U$ in the strong Coupling limit as can be seen in Fig.~\ref{fig_2}(a).
In Fig.~\ref{fig_2}(b) we have plotted $k_BT_C/U$ as a function of the effective
correlation $U/W$. 
This means that the Bloch bandwidth $W$ 
is varied while the Coulomb interaction $U$  is
kept fixed. Note, that this corresponds to the situation in  semi-infinite 
lattices as well as in thin films 
where $U$ is uniform for all layers, the effective bandwidth 
($\sim\mbox{\scriptsize$\sqrt{\mbox{$\Delta_\alpha^{(2)}$}}$}$) 
of the BDOS being, however, different at the surface and in the bulk. 
As we have discussed above, in the strong coupling regime 
the energy scale being
important for the magnetic properties is given by $W$. 
Therefore, for fixed $U$ the Curie temperature within the SDA 
decreases proportional to
$1/W$ as $U/W\rightarrow\infty$ (see Fig.~\ref{fig_2}(b)).
In the limit $W=0$ we find a vanishing Curie temperature, which reproduces the 
exact result in the so-called ``atomic limit'' of vanishing hopping.  
Again, the Hartree-Fock solution fails to describe the strong coupling behavior
correctly, giving a finite Curie temperature even for $W=0$. Since the energy
scale determining the magnetic properties is given by $U$ within the HFA 
the Curie temperature saturates at a finite value as $W\rightarrow0$ in
Fig.~\ref{fig_2}(b).
By applying the results of Fig.~\ref{fig_2}(b)
to the semi-infinite system it is clear  that in the limit of strong Coulomb
interaction  the reduced effective bandwidth in the surface layer 
($\mbox{\scriptsize$\sqrt{\mbox{$\Delta_s^{(2)}$}}$}<
\mbox{\scriptsize$\sqrt{\mbox{$\Delta_b^{(2)}$}}$}$) 
enhances the effective correlation but reduces the magnetic stability at the
surface.
In addition, Fig.~\ref{fig_2}(b) explains the qualitatively different 
behavior concerning the 
magnetic stability at the surface within the  HFA  and the SDA.

We want to stress that both arguments (i) and (ii) that explain the reduced
magnetic stability at the surface in strongly correlated electron systems are
not restricted or linked to the SDA.  Qualitatively, 
the same behavior is to be expected
within all theories that correctly reproduce the exact results of the $W/U$
perturbation theory of Harris and Lange \cite{HL67}. 
It is clear that argument (ii) holds only for strong Coulomb interaction. 
In the case of  moderate coupling strengths the effective correlation $U/W$ 
has to exceed a lower critical bound for spontaneous ferromagnetism to occur.

\begin{figure}
 \centerline{\epsfig{file=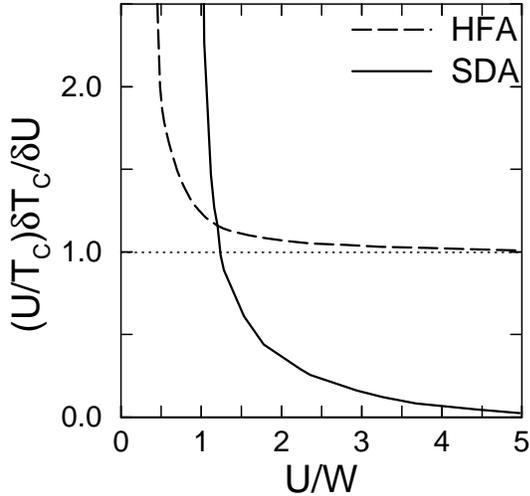,width=0.8\linewidth}}
\caption{$\frac{U}{T_C}(\frac{\partial T_C}{\partial U})$ 
(see Eq.~(\ref{eq:taylor})) calculated within the HFA and the SDA
as a function of the effective correlation $U/W$ for
an fcc lattice (bulk calculation). Further parameter: $n=1.4$.}
 \label{fig_3}
\end{figure}

Let us discuss the boundary between these two regimes (moderate
coupling $\leftrightarrow$ strong coupling behavior)
in more detail. Due to the scaling property of the Hubbard model the following
relation holds \cite{Bemerkung}:
\begin{equation}
T_C(U,rW)=rT_C(\frac{U}{r},W), \qquad r>0\label{eq:scaling}
\end{equation}
Eq.~\ref{eq:scaling} expresses the fact that for constant $U/W$  the energy
scale of the Hubbard model is determined by $W$.  A Taylor expansion of the
right hand side  of (\ref{eq:scaling}) at
$r=1$ up to the first order yields:
\begin{equation}\label{eq:taylor}
T_C(U,rW)=T_C(U,W)\left[ r+\frac{U}{T_C}\left(\frac{\partial T_C}{\partial U}
\right) (1-r)\right]
\end{equation} 
For $\frac{U}{T_C}(\frac{\partial T_C}{\partial U})>1$ a reduction of the
bandwidth $W$ by a factor $r$ will increase the Curie temperature, for 
$\frac{U}{T_C}(\frac{\partial T_C}{\partial U})<1$ the Curie temperature will be
reduced. 
In Fig.~\ref{fig_3} numerical results for 
$\frac{U}{T_C}(\frac{\partial T_C}{\partial U})$ are plotted 
as a function of $U/W$ for both the SDA and the HFA. 
$\frac{U}{T_C}(\frac{\partial T_C}{\partial U})$,  that is defined in the
ferromagnetic phase only, diverges as $U/W$ approaches the lower critical
correlation. Thus, for moderate coupling strength we find 
$\frac{U}{T_C}(\frac{\partial T_C}{\partial U})>1$ within the HFA and the SDA. 
However,
within the SDA $\frac{U}{T_C}(\frac{\partial T_C}{\partial U})$ becomes
smaller than 1 very rapidly with increasing $U/W$
and vanishes as $U/W\rightarrow\infty$. Thus, by inspecting 
$\frac{U}{T_C}(\frac{\partial T_C}{\partial U})$ we can distinguish between two
different regimes: For moderate correlation we expect the magnetic stability at
the surface to be enhanced, whereas it will be reduced for strong Coulomb
interaction. Again, a similar behavior is to be expected within other strong
coupling approaches to the Hubbard model as well.

\begin{figure}[tb]
\centerline{\epsfig{file=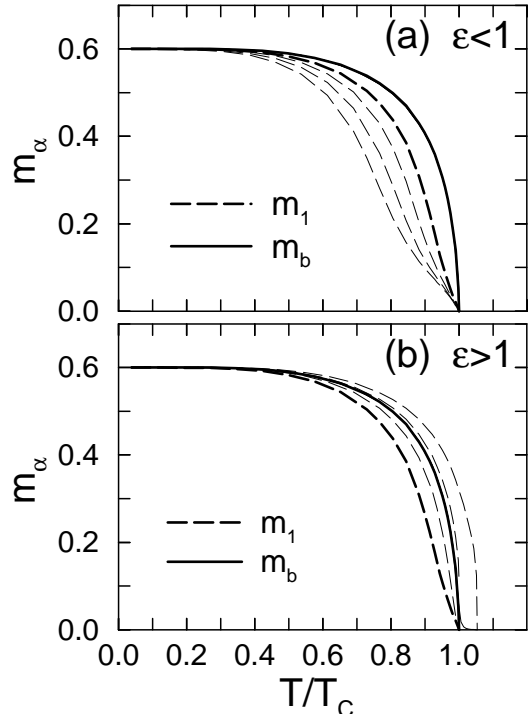,width=0.8\linewidth}}
\caption{Bulk magnetization $m_b$ 
and surface layer magnetization $m_1$
as a function of the reduced temperature $T/T_C$
for a semi-infinite fcc lattice with a (111) surface.
The hopping integral  within the first layer  has been varied according to
$T_{{\bbox k}}^{11}=T_{01}+\epsilon t \gamma_{||}({\bbox k}),\,\,\epsilon>0$.
(a) $\epsilon=1.0,\,\,0.9,\,\,0.8,\,\,0.7$.
(b) $\epsilon=1.0,\,\,1.1,\,\,1.2,\,\,1.3$.
The thick lines correspond to the uniform situation $\epsilon=1.0$. 
Further parameters: $U/W=3$; $n=1.4$} \label{fig_4}
\end{figure}

To further investigate the magnetic stability at the surface we have modified
the hopping integral within the 
surface layer of the semi-infinite fcc(111)
lattice according to 
$T_{{\bbox k}}^{11}=T_{01}+\epsilon t \gamma_{||}({\bbox k})$, $\epsilon>0$.
In Fig.~\ref{fig_4} the corresponding layer magnetizations are shown as a
function of the reduced temperature $T/T_C^{bulk}$. We have 
considered deviations from the unperturbed hopping  $\epsilon=1.0$ 
within 30\% only.
A reduced hopping integral in the surface layer 
($\epsilon<1$, Fig.~\ref{fig_4}(a))
results in a reduced magnetic stability at the surface.
As in the uniform situation ($\epsilon=1$, see also Fig.~\ref{fig_1}) 
we find a unique Curie temperature
$T_C=T_C^{bulk}$  for all $\epsilon<1$ 
due to the coupling between
the surface layer and the bulk. 
As the hopping at the surface is enhanced compared to the bulk
($\epsilon>1$, Fig.~\ref{fig_4}(b)), the magnetic stability at the
surface increases. Note that already a 30\% enhancement of the surface
hopping integral ($\epsilon=1.3$) leads to an enhanced surface Curie 
temperature $T_C^{surf}$ which lies
clearly above $T_C^{bulk}$.  
For $T_C^{bulk}<T<T_C^{surf}$ 
the polarization of the surface layer induces 
a finite  polarization in the adjacent layers. However, 
the induced magnetization  decreases very fast with increasing distance
from the surface. 
In this sense we find two separate Curie temperatures in the system.
The bulk Curie temperature is, of course, not affected by the
enhanced surface Curie temperature. 
Qualitatively, the influence of the modified hopping  at the surface on
the magnetic stability can be understood by argument (ii) that was introduced
above.  A reduced (enhanced) hopping integral at the surface leads to a
reduction (enhancement) of the effective bandwidth of the 
surface layer BDOS and, therefore, to a reduced
(enhanced) magnetic stability. 
Again, the opposite behavior is found within the HFA.

\section{conclusion}
In conclusion we have investigated the magnetic stability at the surface in
strongly correlated electron systems within the semi-infinite Hubbard model. 
For strong Coulomb correlation we find the magnetic stability  at the surface to
be reduced compared to the bulk. This behavior represents 
a pure correlation effect that
cannot be explained within the
well-known Stoner-picture of bandmagnetism which is justified for 
weak Coulomb interaction only. 
However, based on exact results concerning the general
shape of the spectral density in the strong coupling limit the reduced magnetic
stability at the surface can be understood by a moment analysis of the BDOS. It has
been shown that both the reduced variance (effective bandwidth) as well as the reduced
skewness of the surface layer BDOS  tend to reduce the magnetic stability at the
surface.  

By modifying the nearest neighbor hopping integral within the surface layer we have
investigated the  magnetic stability at the surface in more detail. Again, the same
trend is observed.  A reduced (enhanced) hopping integral at the surface leads to a
reduced (enhanced) magnetic stability at the surface. 
For an fcc(111) geometry
already a 30\% enhancement of
the surface hopping integral results in a  surface Curie temperature which lies well
above the bulk Curie temperature.

\acknowledgments{
This work has been done within the Sonderforschungsbereich 290 (``Metallische
d\"{u}nne Filme: Struktur, Magnetismus und elektronische Eigenschaften'') of the
Deutsche Forschungsgemeinschaft.}


\begin{references}
\bibitem{Bemerkung}
This can  be seen, e. g., by
inspecting the grand canonical partition function.
\end{references}
\end{document}